\documentclass[10pt,twoside,a4paper,twocolumn]{article}

\usepackage[utf8]{inputenc}
\usepackage{graphicx}% Include figure files
\usepackage{dcolumn}% Align table columns on decimal point
\usepackage{bm}% bold math
\usepackage{authblk}% For the affiliation management
\usepackage{amssymb}% More advanced math 
\usepackage{amsmath}% Better math
\usepackage{flushend}% Make columns on last page of equal length
\usepackage{placeins}  % For FloatBarrier
\usepackage{url}

\makeatletter
\renewcommand*{\@fnsymbol}[1]{\ensuremath{\ifcase#1\or *\or \dagger\or \ddagger\or
\mathsection\or \mathparagraph\or \|\or **\or \dagger\dagger
\or \ddagger\ddagger \else\@ctrerr\fi}}
\makeatother

\newcommand{\thetaconv}{\ensuremath{\theta_{\text{con}}}}
\newcommand{\thetareso}{\ensuremath{\theta_{\text{res}}}}

\newcommand{\thetaobsv}{\ensuremath{\theta_{\text{obs}}}}
\newcommand{\thetacalc}{\ensuremath{\theta_{\text{cal}}}}
\newcommand{\Nu}{\ensuremath{N_{\text{u}}}}
\newcommand{\Nk}{\ensuremath{N_{\text{k}}}}
\newcommand{\NbCl}{\ensuremath{\text{Nb}_3\text{Cl}_8}}
\newcommand{\MoS}{\ensuremath{\text{Mo}\text{S}_2}}

\newcommand{\myRe}{\ensuremath{\text{Re}}}
\newcommand{\vecr}{\ensuremath{\vec{r}}}
\newcommand{\vecR}{\ensuremath{\vec{R}}}

\newcommand{\bigVecR}{\ensuremath{\vec{\textit{\textbf{R}}}}}

\newcommand{\scE}{\ensuremath{\text{\textsc{e}}}}

\newcommand{\scEm}{\ensuremath{\text{\textsc{e}-}}}
\newcommand{\resp}[1]{#1}  % Build with this one for the print copy
\newcommand{\respout}[1]{}

\begin{document}

\title{Overcoming information reduced data and experimentally uncertain parameters in ptychography with regularized optimization}% Force line breaks with \\

\author[1]{Marcel Schloz}%
\author[1]{Thomas C. Pekin}%
\author[2]{Zhen Chen}%
\author[1]{Wouter Van den Broek\thanks{Correspondence should be addressed to Wouter Van den Broek (vandenbroek@physik.hu-berlin.de)}}%
\author[2,3]{David A. Muller}%
\author[1]{Christoph T. Koch}%

\affil[1]{Institut f{\"u}r Physik \& IRIS Adlershof, Humboldt Universit{\"a}t zu Berlin, 12489 Berlin, Germany}
\affil[2]{School of Applied and Engineering Physics, Cornell University, Ithaca, NY 14853, USA}
\affil[3]{Kavli Institute at Cornell for Nanoscale Science, Ithaca, NY 14853, USA}

\date{\today}

\maketitle

\paragraph{Abstract} The overdetermination of the mathematical problem underlying ptychography is reduced by a host of experimentally more desirable settings.  Furthermore, reconstruction of the sample-induced phase shift is typically limited by uncertainty in the experimental parameters and finite sample thicknesses. 
Presented is a \resp{conjugate gradient descent}
%based
algorithm, regularized optimization for ptychography (ROP), that recovers the partially known experimental parameters along with the phase shift, improves resolution by incorporating the multislice formalism to treat finite sample thicknesses, and includes regularization in the optimization process, thus achieving reliable results from noisy data with severely reduced and underdetermined information.

\section{Introduction}

Ptychography recovers the phase shift of scattered radiation within a probed specimen from a series of diffraction patterns \cite{rodenburg2008ptychography}. It evolved due to improvements in computing power, detector hardware and algorithm development from a technique with relatively limited practical application to a mature technique that is popular in light microscopy \cite{rodenburg2007transmission}, the X-ray community \cite{rodenburg2007hard, holler2017high} and electron microscopy, where it has recently reached sub-angstrom resolution \cite{jiang2018electron}. Its principle is based on illuminating the specimen with a spatially limited probe that is shifted such that an overlap between illuminated areas provides information redundancy. 

One of the main benefits of ptychography is that spatial resolution is not limited by the imaging optics \cite{rodenburg2008ptychography}. However, in practice, it comes at the price of a challenging task for the reconstruction algorithm to solve the phase problem. While the algorithm must recover the phase, the quality of the reconstruction depends strongly on the precise knowledge of the shape \cite{guizar2008phase, thibault2008high, maiden2009improved, kahnt2019coupled}, the coherence \cite{jiang2016optexp}, and the positions \cite{maiden2012annealing, zhang2013translation, tripathi2014ptychographic} of the probe. If these parameters are not initially known, they are required to be retrieved by the algorithm during the reconstruction process.
\resp{Initially, ptychography relied on the projection approximation, which restricted the method to relatively thin samples \cite{rodenburg2008ptychography, thibault2008high}. }
%Conventionally, ptychography relies on the phase object approximation, which restricts the method to relatively thin samples \cite{rodenburg2008ptychography, thibault2008high}. 
To account for multiple scattering in thicker samples, a multislice method can be adopted \cite{van2012method,van2013general, maiden2012ptychographic, tsai2016x, li2018multi, gilles20183d, nikitin2019photon}. In cases where the noise level in the diffraction data is too high and/or the oversampling limit is not reached, ptychography reconstruction algorithms tend to either converge to the wrong solution or are too unstable to reach convergence \cite{thibault2009probe}. 

A variety of ptychographic reconstruction algorithms exist, utilizing either iterative~\cite{maiden2009improved} or direct inversion~\cite{yang2016natcom} to solve for the phase of the object, and sometimes for the probe shape and/or positions~\cite{maiden2012ptychographic}. These iterative algorithms have introduced gradient based optimization~\cite{thibault2012njp} (including the use of finite difference derivatives in the optimization of the beam positions~\cite{dwivedi2018ultram}), multislice reconstructions~\cite{maiden2012ptychographic,van2013general} as well as regularization~\cite{thibault2012njp, katkovnik2013sparse, van2013general}, but these innovations have not been previously combined.

We introduce regularized optimization for ptychography (ROP), which is a multislice-capable 
%derivative-based gradient descent 
\resp{conjugate gradient descent}
reconstruction algorithm that can reconstruct the phase while simultaneously retrieving the probe shape and positions. ROP is an extension to the inversion of dynamical electron scattering (IDES) algorithm, which has been successfully demonstrated to achieve inversion from simulated high resolution transmission electron microscope images, diffraction data and experimental optical Fourier ptychography data \cite{van2012method,van2013general,koch2014crp,jiang2016optexp}. \resp{The reconstruction algorithm adapts the design of an Artificial Neural Network (ANN), a model very well established in the machine learning community, to efficiently calculate the gradients and} the reconstructed phase\resp{, which} can be regularized during optimization to ensure sparseness. 

In this paper it is shown that the conditioning of the reconstruction problem, characterized by the oversampling ratio \cite{miao1998joptsoc}, gets worse for desirable experimental conditions such as smaller observed diffraction patterns and larger step sizes, which can increase recording speed, field of view, or reduce the amount of acquired data, and more convergent electron beams, which improve resolution.  A smaller beam support which improves computation speed also worsens the conditioning.

The power of using ROP is demonstrated by relaxing the experimental requirements and significantly reducing the oversampling ratio to even \resp{below unity, where the problem becomes underdetermined}. It is thus shown that the incoherent resolution can be reached from just the intensities in the central diffraction discs, due to the regularization better conditioning the phase problem.  Furthermore, only partially known experimental conditions such as the beam shape and the beam positions can be treated, as can relatively thick specimens due to the incorporation of the multislice algorithm.  These improvements will enable the application of ptychography at higher frame rates and scanning speeds, more convergent probes, and larger fields of view.  The reduced data load per recording will furthermore facilitate a fast computation of the reconstruction.

The paper is organized as follows.  In Section \ref{sec:method} the image formation in electron ptychography and its impact on the inverse problem are treated.  Furthermore, an overview of the settings for the simulations and the experiments is given. In Section \ref{sec:results} the probe and position correction are demonstrated on simulated $\MoS$ data.  Then, the effect of regularization on underdetermined data is investigated, and reconstructions from experimental $\MoS$ and $\NbCl$ data are presented.  Finally, the conclusions are drawn in Section \ref{sec:conclusions}.

\section{Methods}
\label{sec:method}

\subsection{Image formation in electron ptychography}
\label{sec:imfoelpt}

We consider the multi-slice method for the scattering process of an electron beam that passes through a sample. In this framework, the Schr\"odinger equation has a solution that is expressed for an electron wavefunction that evolves in a specimen which has been sectioned into $Z$ multiple slices. The wavefunction $\psi^{z}(\vec{r})$ is alternatively transmitted through a slice $z$ and propagated to the next slice $z+1$ by the following equation:
\begin{align}
\begin{split}
\label{eq:MultiSlice}
  \psi^{z+1}(\vec{r}) = & \psi^{z}(\vec{r})  \exp\left( i \sigma V^z(\vec{r}) \right) \\
  & \otimes \frac{ -i }{ \lambda \Delta z } \exp \left( \frac{ i \pi }{ \lambda \Delta z } \Vert \vec{r} \Vert_2 \right),
\end{split}
\end{align}
where $\otimes$ denotes the convolution operation, $\sigma$ the interaction constant, $V^z(\vec{r})$ a complex quantity, with its real part $V^z_{re}(\vec{r})$ the projected atomic potential and its imaginary part $V^z_{im}(\vec{r})$ the absorptive potential at slice $z$, $\lambda$ the wavelength, $\Delta z$ the slice thickness, and $\vec{r}$ the lateral real-space vector coordinates, respectively. Since $V^z(\vec{r})$ is complex,  this model accounts for the amplitude contrast. Based on the 
%strong phase object 
\resp{projection} approximation,
the transmission function at slice $z$ can be defined as $t^z = e^{i \sigma V^z(\vec{r})}$, and the convolution kernel in the second half of Eq. (\ref{eq:MultiSlice}) is the so-called Fresnel propagator responsible for free-space transmission to the next slice. The wave exiting the sample $\psi^{Z+1}(\vec{r})$ is further propagated to the detector, which can be modelled by a Fourier transform and the intensity of the diffraction pattern $p$ is thus defined as: 
\begin{eqnarray}
I_p(V, \psi^0, \vecR) = |\mathcal{F}(\psi^{Z+1}(\vec{r}))|^2.
\end{eqnarray}
The multiplication in real space of incoming wave and transmission function amounts to a convolution in reciprocal space, causing wrap-around artifacts due to periodic boundary conditions. Following \cite{kirklandBook}, these artifacts are prevented by setting to zero the frequencies above two-thirds of the Nyquist frequency for each slice in the multislice algorithm, resulting in a diffraction pattern that is zero beyond the circle defined by this limit. For each diffraction pattern $p$ in the stack of the ptychographic dataset, consisting of $P$ diffraction patterns, the incoming wave $\psi^{0}(\vec{r} - \vec{R}_p )$ is shifted by $\vec{R}_p$. 

\begin{figure}
\includegraphics[width=0.99\linewidth]{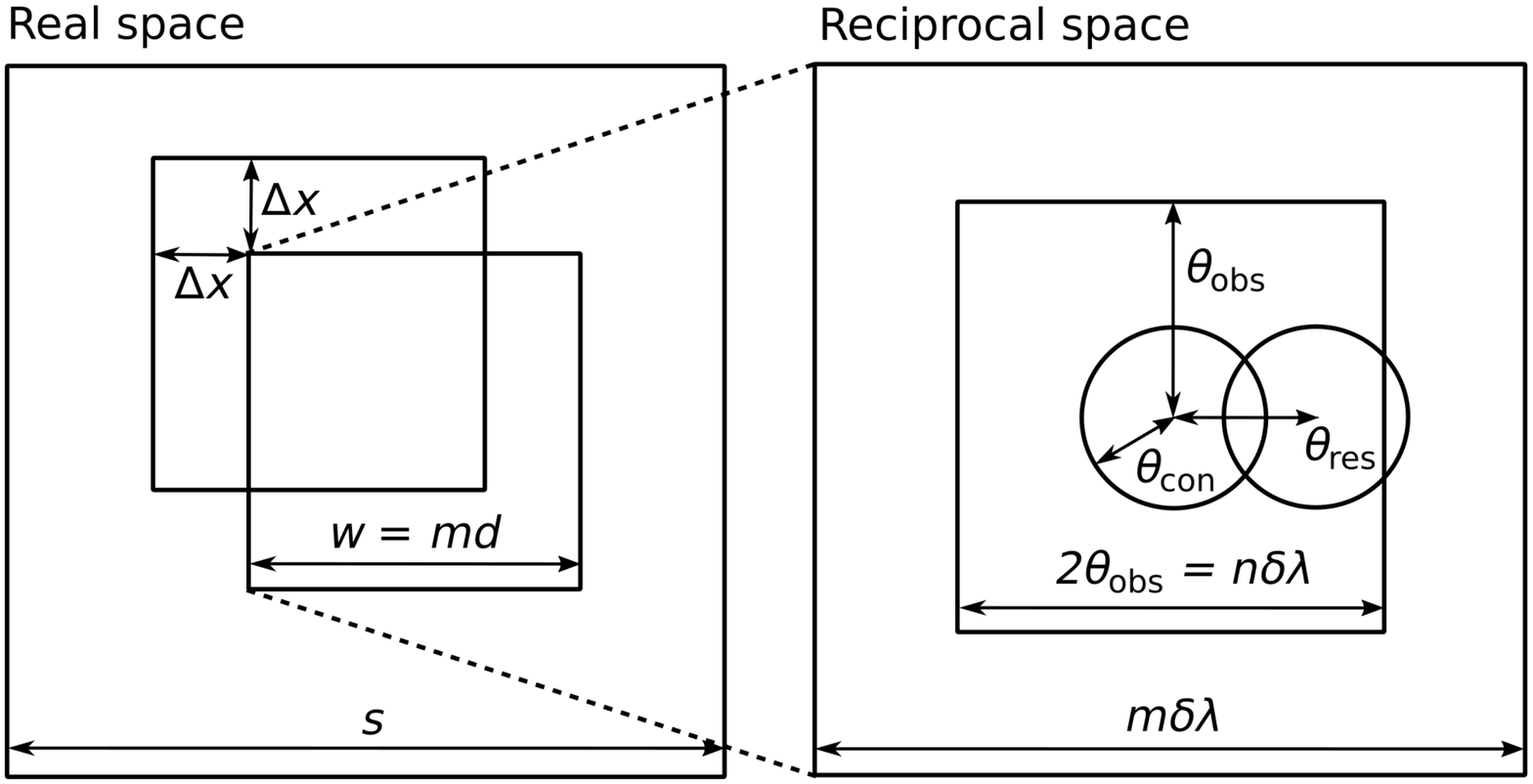}
\centering
\caption{Overview of the parameters of the ptychography set-up, in real and reciprocal space.}
\label{fig:realrecip}
\end{figure}

The main variables of the ptychography set-up are illustrated in Figure~\ref{fig:realrecip}. The beam is scanned with a step size of $\Delta x$, and has a real-space support of width $w$, divided into $m$ pixels of width $d$. This translates into an $m\times m$ diffraction pattern in reciprocal space with pixel size $\delta = 1/w$. The experimentally recorded diffraction patterns are $n \times n$ pixels wide, with $n \leq m$, and the variable $\thetaobsv$ is their width.  Furthermore, $\thetaconv$ stands for the beam's convergence semi-angle.  Reconstruction aims for a resolution $r$, corresponding to the angular frequency $\thetareso$.  As detailed in Appendix \ref{app:underdata}, 
% under the phase object  approximation 
\resp{when single scattering is dominant,}
$\thetareso$ gets smeared out between $\thetareso - \thetaconv$ and $\thetareso + \thetaconv$.  This implies that ptychography is able to achieve the incoherent resolution limit of $2\thetaconv$ from just the intensities within the central disc.

Retrieving the electrostatic object potential $V$ from the measurements is a mathematical problem with a number of knowns, $N_\text{k}$, given by the number of pixels in the measurements, and a number of unknowns, $N_\text{u}$, determined by double the number of pixels in the reconstructed complex potential. The ratio $N_\text{k}/N_\text{u}$ is the so-called oversampling ratio \cite{miao1998joptsoc}, for which in Appendix \ref{app:underdata} an expression is given.  In the limiting case of a large scan area, the incoherent resolution $r$, and thin specimens, the expression reduces to
\begin{eqnarray}
  \frac{N_\text{k}}{N_\text{u}} = \frac{2}{81} \frac{ \thetaobsv^2 }{\thetaconv^2} \frac { w^2 } {\Delta x^2}. \label{eq:NuNkRed}
\end{eqnarray}
It is thus shown that the problem gets conditioned worse for smaller observed diffraction patterns, larger step sizes and smaller beam support, properties that improve both recording and computation speed.  Furthermore, a more convergent electron beam, through a larger opening angle, further reduces the oversampling ratio.

\subsection{The inverse problem}

The following derivations are carried out for a single beam position and the associated diffraction pattern. The index  $p$ is thus dropped for the time being, but will be reintroduced from Eq. (\ref{eq:batchedLoss}) onward to treat the interdependence.  Furthermore, the variable $\sigma$ is absorbed into $V$ for brevity.

The discrepancy between simulation and experiment of a single diffraction pattern is quantified by an error metric $\mathcal{E}$. In this paper, we define three different metrics: 
\begin{align}
\mathcal{E}_1 & = \sum_k^{n \times n} |I_k (V, \psi^0, \vecR) - J_k|, \\
\mathcal{E}_2 &=  \sum_k^{n \times n} (I_k (V, \psi^0, \vecR) - J_k )^2, \\
\mathcal{E}_3 &=  \sum_k^{n \times n} (I_k (V, \psi^0, \vecR) - J_k \ln(I_k (V, \psi^0, \vecR))).
\end{align}
To more clearly distinguish the measurements from the model, they are denoted by $J$.  The metric $\mathcal{E}_3$ describes the log-likelihood of the measurements under a Poissonian noise model.  The error metrics only take into account the central $n \times n$ pixels from the $m \times m$ simulated diffraction patterns.

\resp{Derivative based methods allow a free choice of error metric.  When the noise is purely Poissonian, as is the case with direct detectors \cite{tate2016high}, the log-likelihood $\mathcal{E}_3$ is well-suited, while in \cite{jiang2016optexp}, it was shown how $\mathcal{E}_1 $ was advantageous in the presence of small modelling errors.}

\begin{figure}
\includegraphics[width=0.99\linewidth]{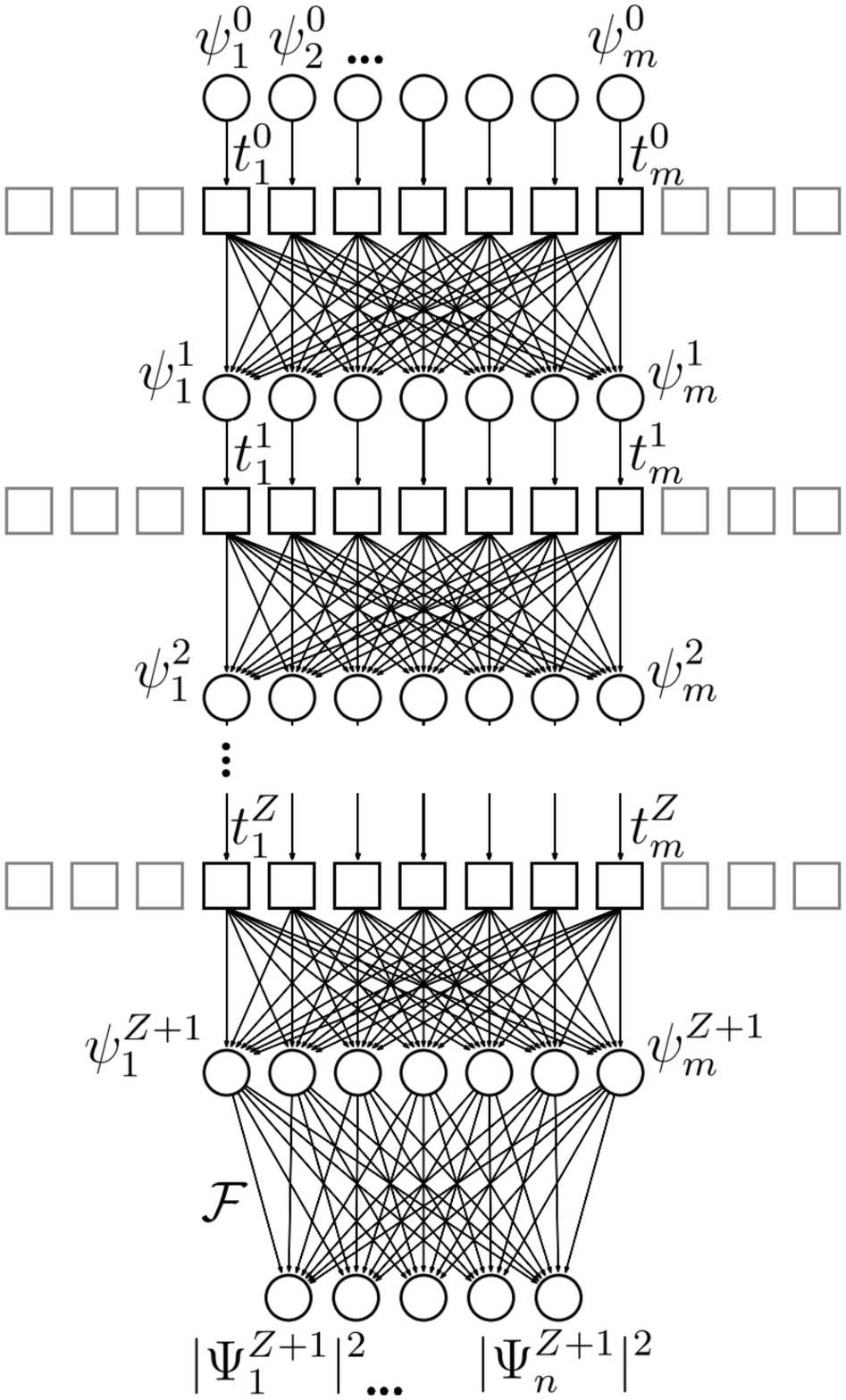}
\centering
\caption{Schematic of the network model. The core ANN is divided into sub-networks by the number of probe positions. For each sub-network, the impinging wave $\psi^z$ is multiplied with a transmission function $t^z$. The next layer of edges and nodes encodes a real-space convolution with the Fresnel propagator, resulting in $\psi^{z+1}$. This is repeated until the exit wave $\psi^{Z+1}$ is produced and the intensity is taken in the far-field.}
\label{fig:network}
\end{figure}

\resp{In addition to the error metric, we introduce a regularization term $\mathcal{R}$ that acts as a sparsity constraint on the object. Among different alternatives~\cite{thibault2012njp, katkovnik2013sparse, van2013general}, we express that atomic resolution images are sparse in the pixel basis and thus define $\mathcal{R}$ by the $\ell_1$-norm of the projected potential:}
\begin{equation}
\label{eq:reg}
\mathcal{R} = \Vert V \Vert_1.
\end{equation}
Optimization methods for inverse problems that involve a regularization need to trade-off the error, corresponding to the exactness of the fit of the model to the given data, against the amount of regularization in the solution. This balancing is controlled by a regularization parameter which in ROP is denoted by $\mu$. Error metric $\mathcal{E}$ and penalization term $\mathcal{R}$ are then combined in a loss function:
\begin{equation}
\ell(V, \psi^0, \vec{R}) = \mathcal{E} + \mu \mathcal{R}.
\end{equation}
 Here, a high $\mu$ value imposes a strong sparsity constraint, while a low $\mu$ value emphasises a strong compliance to the data. Object, probe shape and positions are eventually retrieved by iteratively minimizing the loss function using the respective gradients.  \resp{As illustrated in Figure \ref{fig:network},} \resp{the multislice algorithm is recast as an ANN and we use the backpropagation algorithm to compute the derivatives of the loss w.r.t. the parameter of interest \cite{van2012method,van2013general,koch2014crp,jiang2016optexp} .}

The derivative of the error w.r.t. the atomic potential at each slice is given by:
\begin{align}
\label{eq:DlDV1}
 \frac{\partial \ell(V, \psi^0, \vec{R})}{\partial V^z_{re} } & = -2 \text{Im} \left( \psi^z t^z  \frac{\partial \ell(V, \psi^0, \vec{R})}{\partial \psi^z t^z}  \right), \\
 \frac{\partial \ell(V, \psi^0, \vec{R})}{\partial V^z_{im} } & = -2 \text{Re} \left( \psi^z t^z  \frac{\partial \ell(V, \psi^0, \vec{R})}{\partial \psi^z t^z}  \right),
\end{align}
where $\psi^z t^z$ is computed in the forward propagation and $\frac{\partial \ell(V, \psi^0, \vec{R}_p)}{\partial \psi^z t^z}$ is obtained in the backpropagation algorithm. To account for scattering effects in a thick sample while keeping the complexity of the reconstruction problem low, the atomic potential in all slices can be constrained to be equal, so that the oversampling ratio $N_\text{k}/N_\text{u}$ remains unaffected. Thus, Eq. (\ref{eq:DlDV1}) can be written as:
\begin{equation}
\label{eq:DlDV2}
\frac{\partial \ell(V, \psi^0, \vec{R})}{\partial V} = \frac{1}{Z} \sum^Z_{z = 1} \frac{\partial \ell(V, \psi^0, \vec{R})}{\partial V^z }.
\end{equation}
The derivative of the regularization term $\mathcal{R}$ with respect to the potential is calculated as described in \cite{van2013general}.

The derivative of the loss $\ell$ with respect to the incoming probe, $\psi^0 = \psi^0_{re} + i \psi^0_{im}$, is calculated with a translated object instead of a translated beam to ensure that the result is centered on the origin. Following Appendix \ref{app:derivatives} we have: 
\begin{align}
\begin{split}
 \label{eq:DlDPsiRe}
 & \frac{\partial \ell(V(\vecr+\vec{R}), \psi^0, 0)} {\partial \psi^0_{re}} = \\ &  2 \text{Re} \left(  t^{0} \frac{\partial \ell(V(\vecr+\vec{R}), \psi^0, 0)}{\partial \psi^0 t^0}  \right), 
\end{split}\\
\begin{split}
 \label{eq:DlDPsiIm}
& \frac{\partial \ell(V(\vecr + \vec{R}), \psi^0, 0)}{\partial \psi^0_{im}}   = \\ &  -2 \text{Im} \left( t^{0} \frac{\partial \ell(V( \vecr + \vec{R} ), \psi^0, 0)}{\partial \psi^0 t^0} \right).
\end{split}
\end{align}

The derivative of the error w.r.t. the probe position, also calculated in Appendix \ref{app:derivatives} is:
\begin{align}
\begin{split}
& \frac{\partial \ell (V, \psi^0, \vec{R})}{\partial \vec{R}}  = \\ &  - 2 \myRe \Bigg( \sum_k \frac{\partial \ell (V, \psi^0, \vec{R}) }{\partial \psi_k^0 (\vec{r}-\vec{R})} \left[ \frac{ \partial \psi^0 (\vec{r}-\vec{R})} {\partial \vec{r}} \right]_k  \Bigg). 
\label{eq:DlDR}
\end{split}
\end{align}
The first factor in the summation is the derivative of the loss function with respect to the probe in position $\vec{R}$ and follows directly from the backpropagation; the second factor is the derivative of the probe in position $\vec{R}$ with respect to the spatial coordinate $\vec{r}$ and can be calculated numerically by a convolution with a one dimensional derivative filter in each of the two spatial directions; the index $k$ runs over all elements of the probe and its derivatives. Note that contrary to~\cite{dwivedi2018ultram}, this is expression is analytical.

The forward propagation of the electron wave and the subsequent backpropagation of the loss are computed independently for each diffraction pattern $p$. The updates for atomic potential $V$ and incoming beam $\psi^0$ are then performed with a loss function and gradients that are respectively formed from the entire batch of diffraction patterns $P$,
\begin{align}
\mathcal{L}(V, \psi^0, \vec{\textit{\textbf{R}}})  & = \frac{1}{P}\sum^P_{p=1} \ell(V, \psi^0, \vec{R}_p), \label{eq:batchedLoss}\\
\nabla_V \mathcal{L}(V, \psi^0, \vec{\textit{\textbf{R}}}) & = \frac{1}{P}\sum^P_{p=1} \nabla_V \ell(V, \psi^0, \vec{R}_p), \\
\nabla_{\psi^0} \mathcal{L}(V, \psi^0, \vec{\textit{\textbf{R}}}) & = \frac{1}{P}\sum^P_{p=1} \nabla_{\psi^0} \ell(V, \psi^0, \vec{R}_p),\\
\left[ \nabla_{\vec{\textit{\textbf{R}}}} \mathcal{L}(V, \psi^0, \vec{\textit{\textbf{R}}}) \right]_p & = \frac{1}{P} \nabla_{\vec{R}_p} \ell(V, \psi^0, \vec{R}_p).
\end{align}
It is assumed that the incoming beam $\psi^0$ is constant over the beam positions. Furthermore $\vec{\textit{\textbf{R}}}$ collects the individual vectors $\vec{R}_p$.

The projected potential of the object, the probe shape and the probe positions are retrieved through a non-linear conjugate-gradient method, making use of Polak-Ribi\`ere search-directions $d$ that depend on the batched gradients and the search-directions of the previous iteration~\cite{polak1969note,polyak1969conjugate,nocedalBook}:
\begin{align}
\begin{split}
V_{f+1}^a & = V_f^a \\ - & \alpha_f d_f \left( \nabla_V \mathcal{L}(V_f^a, \psi^{0,a}, \vec{\textit{\textbf{R}}}^a), d_{f-1} \right), \label{eq:dEdV}
\end{split}\\
\begin{split}
\psi^{0,a}_{g+1} &= \psi^{0,a}_g \\- &\beta_g d_g \left(\nabla_{\psi^0} \mathcal{L} (V^{a+1}, \psi^{0,a}_g, \vec{\textit{\textbf{R}}}^a), d_{g-1} \right), \label{eq:dEdPsi}
\end{split}\\
\begin{split}
\vec{\textit{\textbf{R}}}^a_{h+1} &= \vec{\textit{\textbf{R}}}^a_h\\ - & \gamma_{h} d_h \left( \nabla_{\vec{\textit{\textbf{R}}}} \mathcal{L} (V^{a+1}, \psi^{0,a+1}, \bigVecR^a_h), d_{h-1} \right). \label{eq:dEdR}
\end{split}
\end{align}
In each epoch, enumerated by the superscript $a$, the three quantities are optimized iteratively during a so-called sub-epoch.  At the end of the iterations for $V$, enumerated by the subscript $0 \leq f < F$, it holds that $V^{a+1} = V^a_F$, and, \textit{mutatis mutandis}, the same rules hold for the other two quantities.  The step sizes of the respective optimizations $\alpha$, $\beta$ and $\gamma$ are found by using a cubic interpolation. The initial step sizes must be chosen sufficiently small such that only the search space of the local minimum is considered.

%=========================================================

\subsection{Settings for simulated and experimental data}
\label{seq:settings}
To investigate the performance of the algorithm and the regularization effects on the reconstruction, we conducted one simulation and one experiment on both a $\MoS$ and a $\NbCl$ specimen. The first simulation tested for successful convergence of the algorithm when either probe or positions were distorted and the result was supported by an experiment. For the simulation, the ptychographic data of a $\MoS$ monolayer was generated using the code described in \cite{vandenbroek2015fdes} and with parameters that were mostly chosen to match those used in the experiment \cite{jiang2018electron}. \resp{This data, as well as all following simulated data, can be found in Dataset 1~\cite{dataset}}. This experiment, also investigating a $\MoS$ monolayer, was performed on the Titan Themis 300 microscope, operated with an acceleration voltage of $80$~kV ($\lambda = 4.18$~pm) and a convergence semi-angle $\thetaconv$ of $21.4$~mrad. The EMPAD direct electron detector \cite{tate2016high} recorded $124 \times 124$ pixel diffraction patterns and while a $87 \times 51$ probe position scan with a step size $\Delta x$ of $0.021$~nm was used in the experiment, a scan of $100 \times 100$ probe positions was chosen in the simulation.

Our second simulation and experiment aimed at improving the resolution of the reconstruction for limited data from a $\NbCl$ specimen.  A resolution corresponding to twice the beam convergence semi-angle, $\thetareso = 2\thetaconv$, was aimed for.  During reconstruction the angular frequencies were calculated out to $\thetacalc = 3\thetaconv$.  In this case, a measure of spatial resolution is the minimum resolvable distance between two atoms. 

$\NbCl$ is a trigonal system with lattice parameters $a = b = 6.831$~\AA \ and $c = 13.75$~\AA. In the $[001]$ orientation Nb-dumbbells with an atom spacing of $0.67$~\AA \ can be observed. This is shown in Fig. \ref{fig:NbCl_Simulation}(c). At an acceleration voltage of $120$~kV, equivalent to $\lambda = 3.35$~pm, the dumbbell spacing corresponds to a scattering angle of $50$~mrad. 

We simulated ptychographic data using the code described in \cite{vandenbroek2015fdes}, with a semi-convergence angle, $\thetaconv$, set to $28$~mrad so that when $\rho = 2$ the angular frequency of the dumbbell falls well below $\thetareso = \rho\thetaconv = 56$~mrad.  The number of pixels in the beam support of the simulation was chosen to be four times higher than needed for the reconstruction to avoid the so-called ``inverse crime'' fallacy \cite{colton1992}. Because this results in a four-times smaller pixel size in reciprocal space, a four-by-four binning of the simulated diffraction patterns was done before reconstruction. The number of slices in the multislice simulation was set to $35$ with a thickness of $0.1338$~nm per slice, thus covering the crystal's vertical extent of $3c = 4.125$~nm. The scan pattern followed a Halton sequence to avoid artifacts in the reconstruction that could originate from the translation symmetry of a grid scan pattern which has been dubbed ``raster grid pathology'' \cite{thibault2009probe}. A total of 6,600 points within a disc of diameter $4.19$~nm were probed, resulting in an average distance between neighboring beam positions of $\Delta x = 0.0457$~nm. 

The desired resolution corresponded to $2\thetaconv = 56$~mrad, which is therefore sufficient to resolve the dumbbells at $50$~mrad.  The diffraction patterns only contained the central disc, i.e. $\thetaobsv = \thetaconv$. The binning resulted in a beam support of $w = 0.232$~nm wide and diffraction patterns of $n = 4$ pixels, and $m = 20$. The resulting oversampling ratio was $0.47$, or underdetermined by a factor of over two. 

In addition, we generated simulated data similar to the aforementioned data, differing only in the number of beam positions by a factor of four, totalling 26,400 positions.  The diffraction patterns were subsequently binned down by a factor of two, resulting in an oversampling ratio of $6.6$.  The reconstruction from this strongly overdetermined and noiseless data acted as ground truth in the evaluation of the other reconstructions.  The settings of the simulation and reconstruction are detailed in Table~\ref{tab:setupNbCl}. 

For the experiment, a $4$~nm thick $\NbCl$-crystal was produced through exfoliation and transfer on a SiN TEM grid with holes.  It was investigated on the Titan Themis 300 microscope, with an acceleration voltage of $120$~kV ($\lambda = 3.35$~pm) and a convergence semi-angle $\thetaconv$ of $24$~mrad. $124 \times 124$ pixel diffraction patterns were recorded on the EMPAD direct electron detector, with a $64 \times 64$ scan and a scan step size $\Delta x$ of $0.043$~nm.

From these raw data, three reconstructions were calculated, the first from the unaltered data, and then two from reduced data sets. For the first reduced data set (the second reconstruction) the diffraction patterns were binned $3 \times 3$, and only the central $12 \times 12$ pixels were retained, thus yielding diffraction patterns out to $\thetaobsv = 36.8$~mrad, and a beam support of $w = 0.571$~nm. Since $\thetaobsv + \thetaconv = 60.8$~mrad encompasses the dumbbell signal at $50$~mrad, resolved dumbbells were expected.  The second reduced data set (the third reconstruction) is obtained by a $6 \times 6$ binning, retaining only the central $14 \times 14$ pixels. This yielded a beam support of $w = 0.285$~nm and diffraction patterns out to $\thetaobsv = 82.3$~mrad, well beyond the farthest extent of the dumbbell signal at $50$~mrad $+ \thetaconv = 74$~mrad, and hence resolved dumbbells were expected.  The three data sets had an oversampling ratio of $70$, $5.6$ and $7.5$, respectively.  All settings are summarized in Table~\ref{tab:setupNbCl}.

\section{Results}
\label{sec:results}

\subsection{Probe and position correction on simulated M\MakeLowercase{o}S$_2$ data}
\label{sec:simMoS2}
We investigated the performance of the update functions (\ref{eq:dEdPsi}) and (\ref{eq:dEdR}) on the simulated M\MakeLowercase{o}S$_2$ data described in \ref{seq:settings}. For all the tests, optimization was done for 400 iterations, regularization has been neglected, the initial atomic potential was assumed to be vacuum and since a single layer M\MakeLowercase{o}S$_2$  sample exhibits weak phase difference, the depth of the ROP model has been set to one slice.

\begin{figure}
\includegraphics[width=0.99\linewidth]{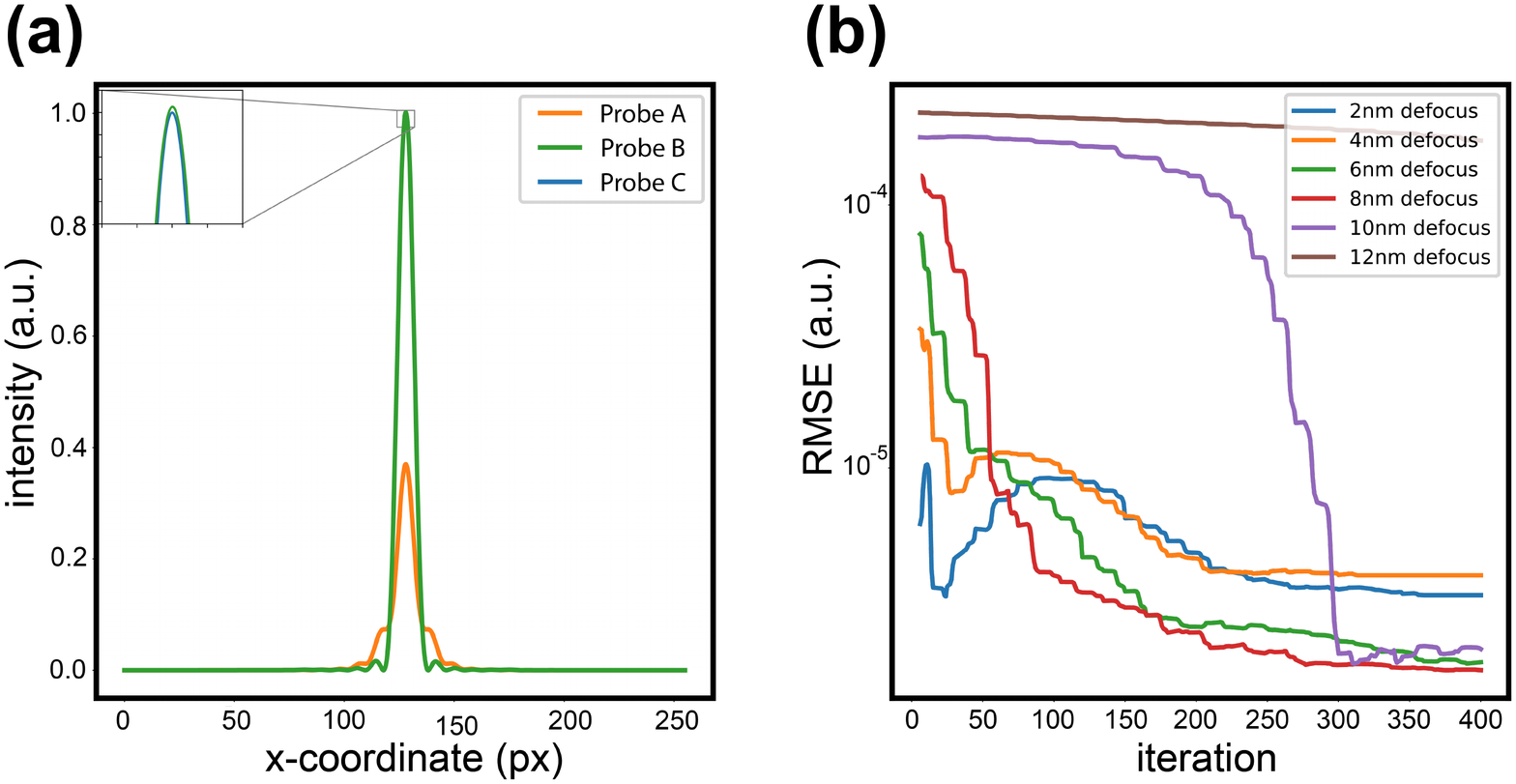}
\centering
\caption{\textbf{a)} Comparison of the intensity profile between the correct incoming probe, a false probe (i.e. defocused by 10nm) and the updated probe. \textbf{b)} RMSE between the correct incoming probe and differently strong deviated incoming probes determined by different defocus values.}
\label{fig:ProbeUpdate}
\end{figure}

For the analysis of the probe shape optimization, a false incoming probe that had been deviated in defocus, relative to the correct probe, was read in. In Figure~\ref{fig:ProbeUpdate}, we apply the optimization for incoming probes that deviate in defocus by up to $12$~nm. The correct probe positions were used, leaving the algorithm to alternate only between the object and probe shape update function. Comparing different optimization strategies, a sub-epoch of $5$ iterations for the object update function and $10$ iterations for the probe shape update function achieved the lowest root mean squared error (RMSE) between the probe shape of the updated probe and the correct probe. We further chose the error metric $\mathcal{E}_1$ and the initial step size $\alpha_0$ for the object update function was set to 1, while the initial step size $\beta_0$ for the probe shape update function was set to 1\scE-5. Using the aforementioned parameter settings, Figure~\ref{fig:ProbeUpdate}(a), shows the intensity line profile of a probe with a defocus of $10$~nm that was taken as the initial guess (Probe A), the updated final probe (Probe B) and the correct probe that was used to generate the data (Probe C). The RMSE of the probe shape between Probe B and Probe C was $2.05$\scEm$6$. Figure~\ref{fig:ProbeUpdate}(b) shows for all the test cases the RMSE of the probe shape between the probe at each iteration of the optimization process and Probe C.  

\begin{figure}
\includegraphics[width=0.99\linewidth]{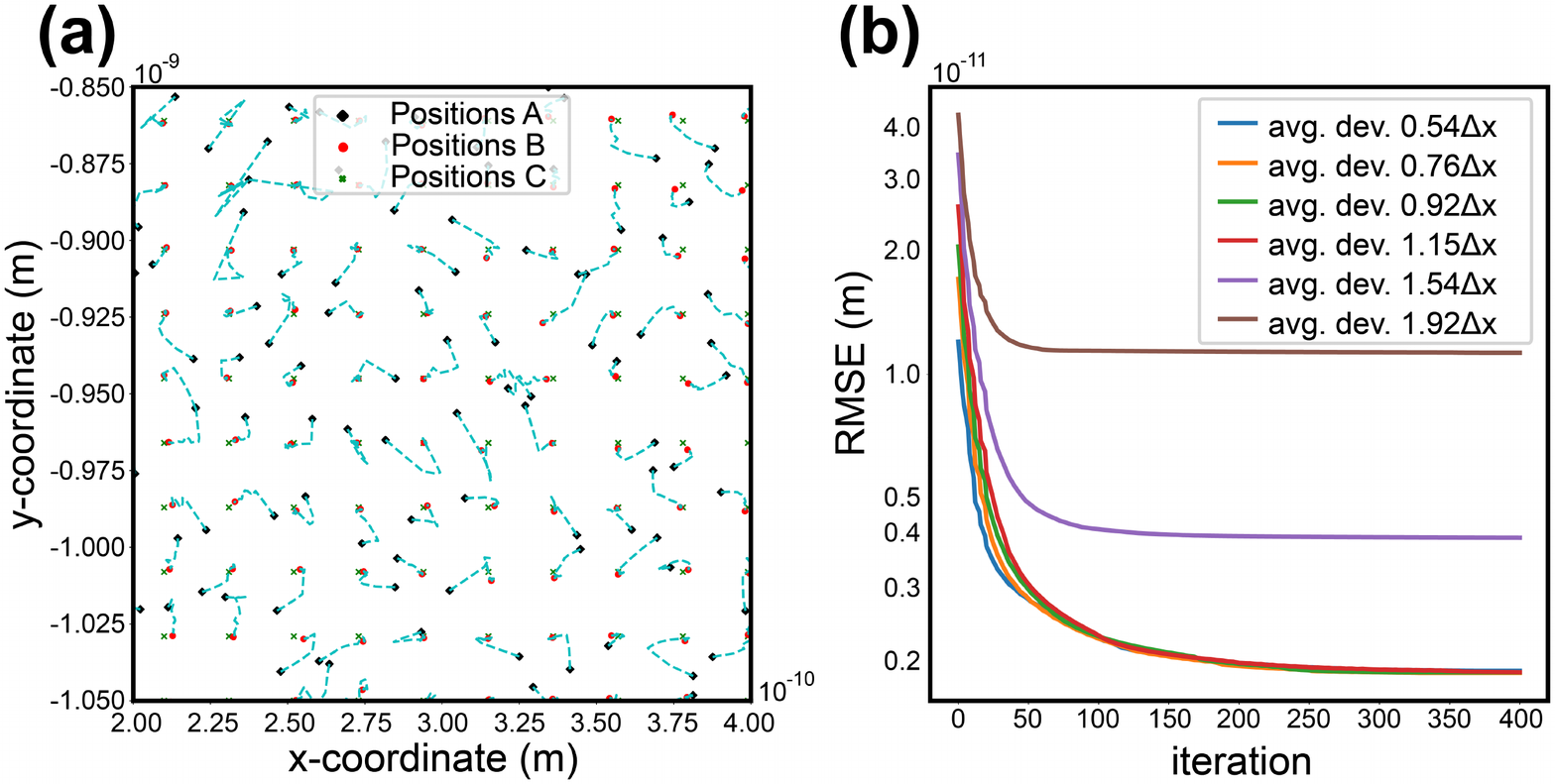}
\centering
\caption{\textbf{a)} Update map of probe positions that have been deviated on average by $0.54 \Delta x$ (with $\Delta x = 0.021\text{nm}$) as they converge to the true probe positions. \textbf{b)} RMSE between the correct probe positions and differently strong randomly deviated probe positions.}
\label{fig:pos_update}
\end{figure}

In our second analysis, we focused on the probe position optimization. Here, the reconstruction algorithm started the optimization with probe positions (Positions A) that were randomly deviated from the correct positions (Positions C) and finally converged to the updated probe positions (Positions B). The performance was then evaluated by taking the RMSE between the Positions at every iteration and Positions C. In this analysis, the correct probe shape was used and therefore only update functions (\ref{eq:dEdV}) and (\ref{eq:dEdR}) were alternated. Sub-epochs of $3$ iterations for the object update function and $4$ for the probe position update function gave the best result. The initial step size $\alpha_0$ for the object update function was set to 1\scE3 and the initial step size $\gamma_0$ for the probe position update function was set to 1\scE5. The optimal error metric appeared to be $\mathcal{E}_2$, showing a much lower final RMSE than $\mathcal{E}_1$. Figure~\ref{fig:pos_update}(a) shows the Positions A for a subsection of the scan grid and an averaged deviation of $0.54\Delta x$. They are compared to the Positions B and the Positions C. The trajectories they form during the optimization process are indicated by a dashed blue line. In Figure~\ref{fig:pos_update}(b), we show the optimization for cases with initial mean deviation starting from $0.54 \Delta x $ up to $1.92 \Delta x$ and compared the RMSE during the optimization process. 

\subsection{Effect of regularization on under-determined data}
\label{sec:simNbCl}
In this section, we analyzed the influence of the regularization $\mathcal{R}$ on the reconstructed object potential. In particular, we concentrated on the simulated $\NbCl$ data set, from which we created $7$ data sets that differed by the electron count per pixel in the central disc, further denoted as electron count. We increased the electron count from $100$ to $1000$ and restricted the analysis to only update function (\ref{eq:dEdV}). The optimization has been limited to 100 iterations and again the initial atomic potential was chosen to be vacuum. The thickness of the $\NbCl$ specimen has been covered in our model by using $7$ slices, each separated by $0.669$~nm and equally updated according to Eq. (\ref{eq:DlDV2}). The initial step size $\alpha_0$ for the object update function was set to 1 and we selected the error metric $\mathcal{E}_2$. 

\begin{figure*}
\includegraphics[width=0.85\textwidth]{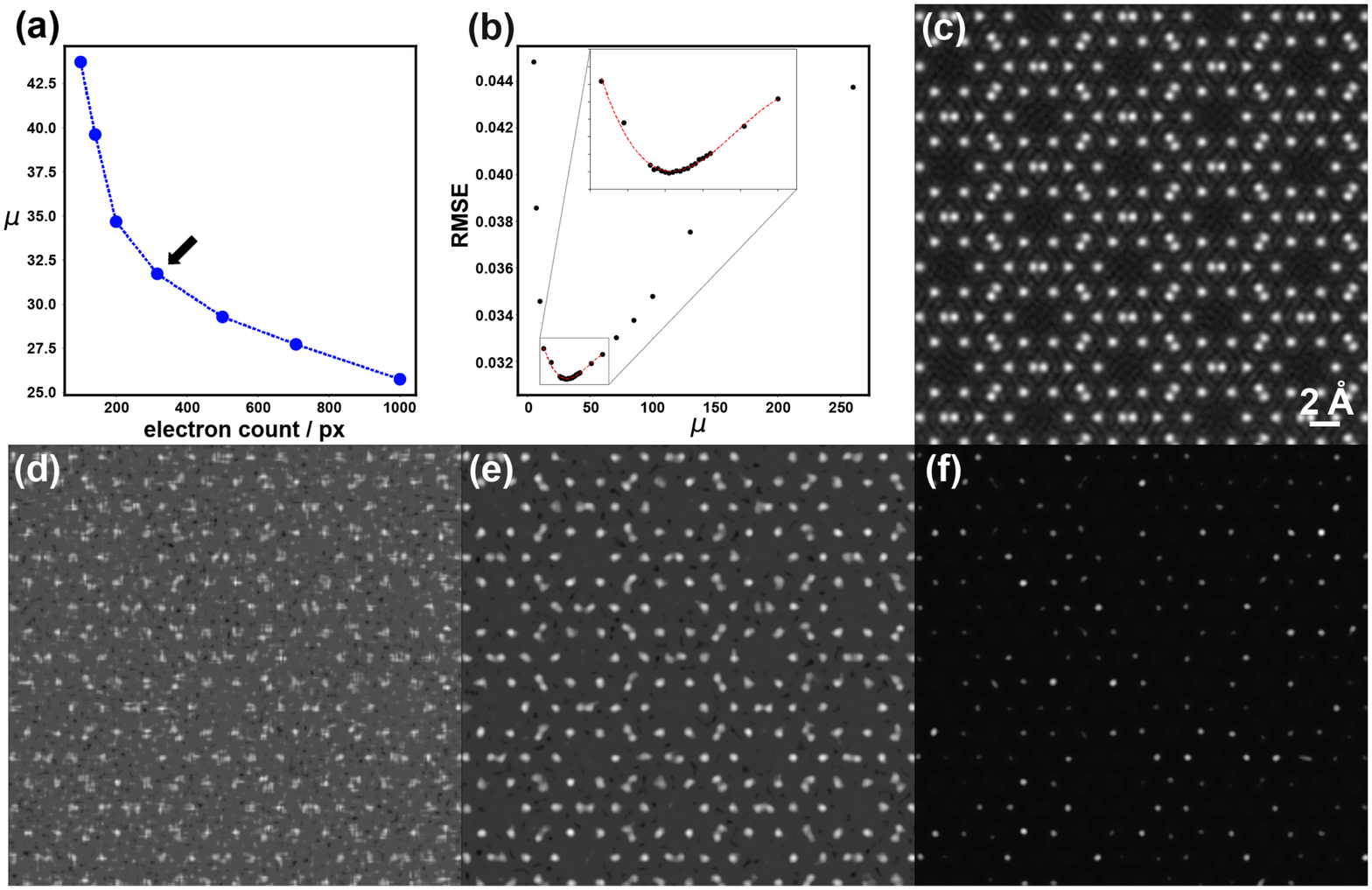}
\centering
\caption{Evaluation of the optimal regularization for (a) the $7$ simulated data sets with different noise levels of a Nb$_3$Cl$_8$ specimen, based on (b) the RMSE between the respective regularized reconstruction and (c) the ground truth reconstruction. For the case of an electron count of $316$, marked with an arrow in (a), the object potential is reconstructed d) under-regularized with $\mu = 5.18$, e) optimally regularized with $\mu = 3.20$\scE${1}$ and f) over-regularized with $\mu = 2.68$\scE${2}$.}
\label{fig:NbCl_Simulation}
\end{figure*}

In Figure~\ref{fig:NbCl_Simulation} we demonstrate the influence of the regularization on the reconstruction quality with a particular focus on the data set that has an electron count of $316$. Figure~\ref{fig:NbCl_Simulation} (a) shows that for each of the $7$ data sets, a different regularization parameter $\mu$ is required. The higher the electron count in the data, i.e. the less noise present, the less regularization is needed for an optimal reconstruction. Here, we estimated the optimal $\mu$ based on the RMSE between the final reconstruction of each test case and a reconstruction from the over-determined data set, acting as a ground truth. Tested regularization parameters covered a range from $5.18$ to $2.68$\scE${2}$ and the optimal $\mu$ was determined as the closest sampling point to the local minimum of a 3rd-order polynomial function that was fitted to a sub-region of the test range. To increase the accuracy of our fit function, we sampled more densely the range close to the local minimum. Figure~\ref{fig:NbCl_Simulation}(b) exemplifies the parameter estimation based on the RMSE for the data set with an electron counts of $316$. The parameter resulting in the lowest RMSE was found to be $\mu = 3.20$\scE${1}$. Figure~\ref{fig:NbCl_Simulation}(c) shows the reconstruction from the ground truth data that is compared to the reconstructions of the test cases. Figure~\ref{fig:NbCl_Simulation}(d) shows a strongly under-regularized reconstruction, corresponding to  $\mu = 5.18$, Figure~\ref{fig:NbCl_Simulation}(e) shows an optimally regularized reconstruction, corresponding to $\mu = 3.20$\scE$1$ and Figure~\ref{fig:NbCl_Simulation}(f) shows a strongly over-regularized reconstruction, corresponding to $\mu = 2.68$\scE${2}$.

\resp{Although $\mu$ was determined through a sweep, other approaches have been considered in \cite{vandenberg2008siam,vandenbroek2019ieee}.}

\subsection{Reconstructions from experimental M\MakeLowercase{o}S$_2$ and N\MakeLowercase{b}$_3$C\MakeLowercase{l}$_8$ data}
\label{sec:experiment}
\begin{figure*}
    \centering
    \includegraphics[width=0.75\textwidth]{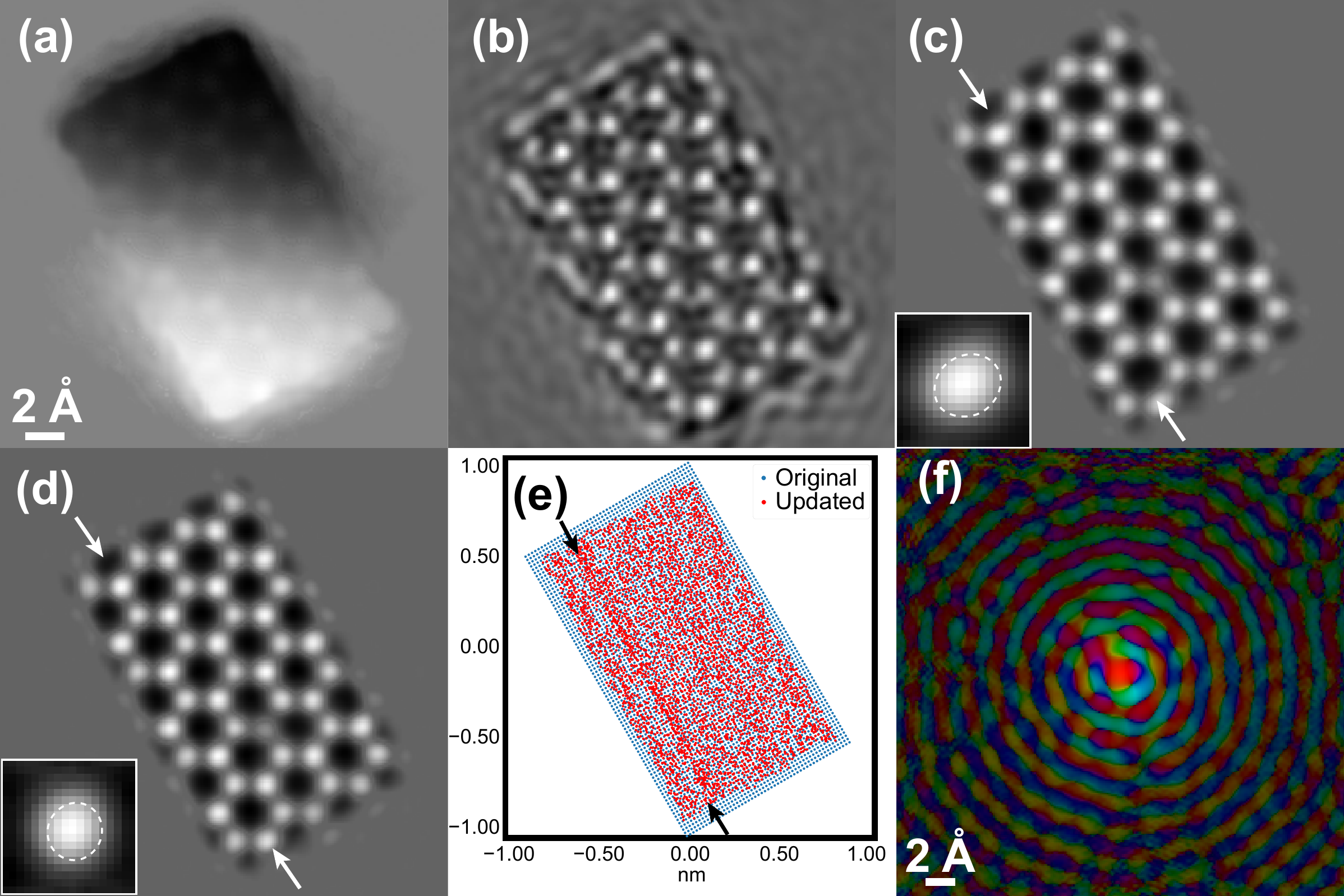}
    \caption{MoS$_2$ object potential after (a) only updating the object, (b) updating the object and probe from an initial default, (c) updating the object and probe, starting with an optimized probe, but with the original grid of probe positions, and (d) with updated probe positions. Notice the elongation of the atoms between the arrows \resp{in (c), but not in (d). Inset in both (c) and (d) are the class averages of the atoms between the white arrows, overlaid with the best fit ellipses to the atoms' half maximum intensities}. (e) The updated probe positions \resp{with black arrows corresponding to the white arrows in (c) and (d)}, and (f) the updated probe used for the final reconstruction, where amplitude to the 1/4 power is represented by intensity, and phase is represented by the HSV colorwheel.}
    \label{fig:mos2_experiment}
\end{figure*}

Figure~\ref{fig:mos2_experiment} shows the result from the \MoS, focusing on the different parts of the algorithm. The settings are shown in Tables~\ref{tab:setupNbCl} and~\ref{tab:setupmosexp}, where $\mu$ is the regularization parameter, $\alpha_0$ is the initial object update step size, $\beta_0$ is the initial probe update step size, and sub ep.~$\alpha$ and sub ep.~$\beta$ are the sub epoch lengths of each in iterations, respectively. Figure~\ref{fig:mos2_experiment}(a) shows the resulting object when only the object is updated, and the rest of the update-able options (probe positions and probe shape) are left at their defaults. This results in an object with a large background ramp of illumination from top to bottom, which mostly obscures the atomic potentials. However, fitting and removing this background plane of potential does reveal underlying atomic potentials.  A possible reason for the phase ramp in the background is a sub-pixel error in positioning the diffraction patterns~\cite{guizar2011phase}.  \resp{Figure}~\ref{fig:mos2_experiment}(b) shows the result of then adding just the probe update, which helps to remove the ramp of illumination, but does not accurately reconstruct the expected spherically symmetric atomic potentials. \resp{Figure}~\ref{fig:mos2_experiment}(c) shows the result when a previously optimized probe, from \resp{\ref{fig:mos2_experiment}}(b), is passed in as an initial argument to the reconstruction, resulting in much more symmetric atomic potentials. Figure~\ref{fig:mos2_experiment}(d) shows the same as \resp{\ref{fig:mos2_experiment}}(c), but now also allowing for the positions to be updated. This results in even more circularly symmetric atoms, particularly in the row between the arrows in \resp{\ref{fig:mos2_experiment}}(c), in which the atoms are elongated. \resp{The change in atom shape can be seen in the subpanels for \ref{fig:mos2_experiment}(c) and \ref{fig:mos2_experiment}(d), in which the average shape of the atom in the rows marked by the white arrows is shown, along with the best fit ellipse to the half maximum intensities of the class average. A clear difference in ellipse major and minor axis length can be seen in \ref{fig:mos2_experiment}(c), with a much less noticeable difference in \ref{fig:mos2_experiment}(d).} The updated positions are shown in \resp{\ref{fig:mos2_experiment}}(e), showing the ability to compensate for \resp{global inaccuracies in probe position, due to } imperfect rotation and pixel size scaling calibrations, as well as \resp{local inaccuracies} in positions in the region \resp{corresponding to the region} between arrows in \resp{\ref{fig:mos2_experiment}}(c) \resp{with} the elongated atoms. \resp{The corresponding region to the area between the white arrows in \ref{fig:mos2_experiment}(c) and \ref{fig:mos2_experiment}(d) is between the black arrows in \ref{fig:mos2_experiment}(e), where a clear shift in positions can be seen, possibly due to microscope instabilities}. Figure~\ref{fig:mos2_experiment}(f) finally shows the reconstructed probe at the sample's exit used for \resp{\ref{fig:mos2_experiment}}(c) and \resp{\ref{fig:mos2_experiment}}(d). Here we show that only through the optimization of probe positions, probe shape, and object itself do we obtain the highest quality reconstructions using ROP.

\begin{figure*}
    \centering
    \includegraphics[width=0.75\textwidth]{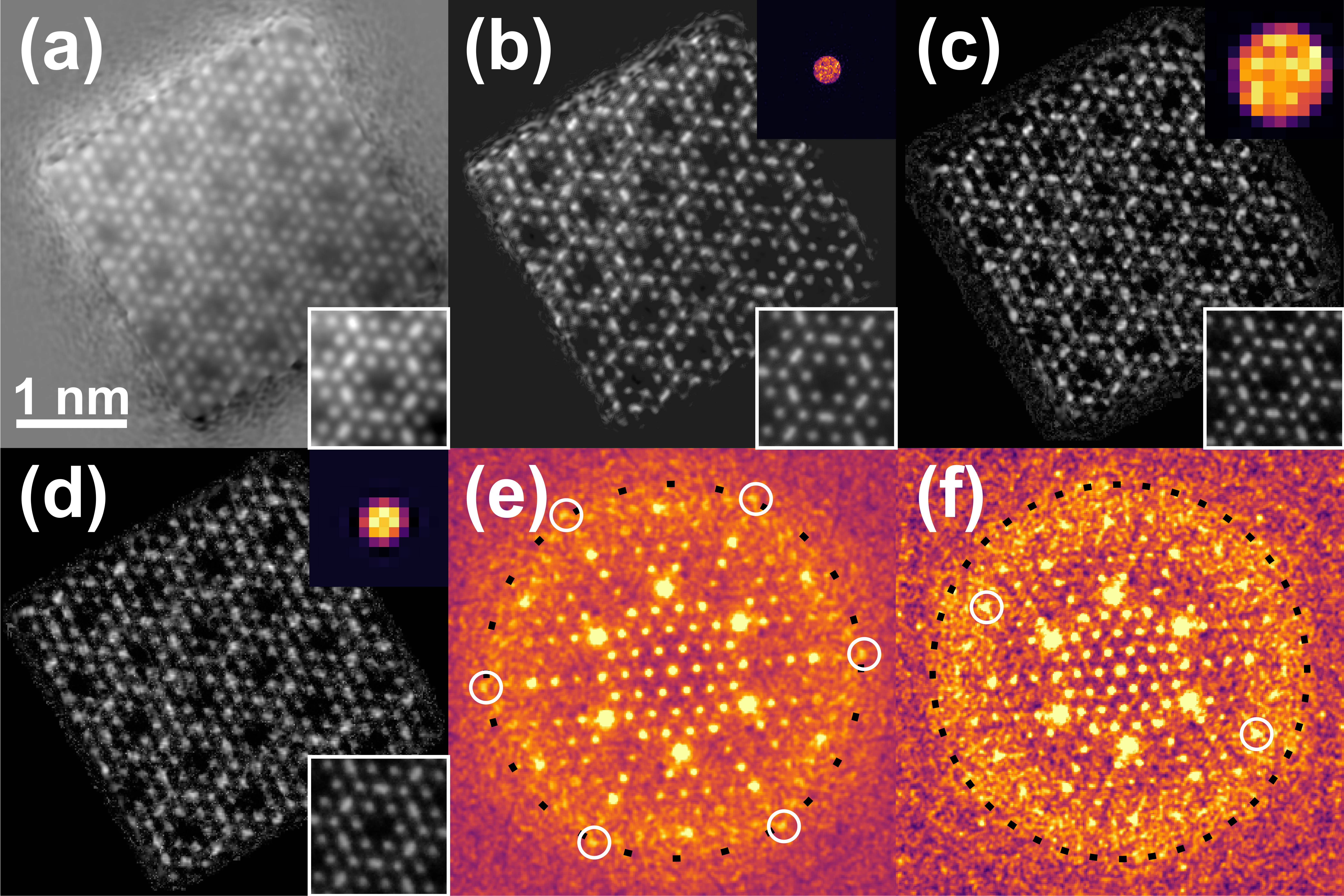}
    \caption{Nb$_3$Cl$_8$ object potential updating the probe and object with (a) the full data set, and one slice, (c) a 3$\times$3 binned data set, with the center 12$\times$12 pixels cropped and a six slice multislice reconstruction, (b) the full dataset with a six slice multislice reconstruction,  and (d) binned by 6, with center 14$\times$14 pixels cropped, also with six slices. Inset in the top right corner are representative diffraction patterns, and in the bottom right corner, class averages of the objects' unit cell. (e) is the Fourier Transform of (c), with the dashed black circle drawn corresponding to the dumbbell resolution of 0.67~\AA, and (f) is the Fourier Transform of (d), with white circles drawn around two representative spots corresponding to a 0.84~\AA{} resolution.}
    \label{fig:nbcl_experiment}
\end{figure*}

Figure~\ref{fig:nbcl_experiment} shows the result from using ROP on data acquired from a 4~nm \resp{slab} of \NbCl, and the benefits of using the multislice approach. The parameters used can be seen in Tables~\ref{tab:setupNbCl} and~\ref{tab:setupnbclexp}, where $\Delta z_{\text{slice}}$ is the propagation distance between slices. In all cases, updated probe positions were used from a prior reconstruction in order to focus on the multislice capabilities of the algorithm. Figure~\ref{fig:nbcl_experiment}(a) shows the result of reconstructing the 
\respout{phase} 
object with only one slice in the beam direction. This ignores all aspects of multiple scattering, and results in the dumbbells being unresolved. In the unit cell class average (lower right), the location of the dumbbells are resolved, but not the spacing between the atoms. \resp{Figure} \ref{fig:nbcl_experiment}(\resp{b}) shows the same result, but this time having 6 slices, with a fixed propagation distance of 0.66875~nm, \resp{corresponding to half the vertical lattice parameter}, thereby matching the sample's total thickness of 4~nm. The dumbbells are sharper, resolvable in individual cases, and can be seen in the class average. Figure~\ref{fig:nbcl_experiment}(\resp{c}) and \resp{\ref{fig:nbcl_experiment}}(\resp{d}) both show data reduction techniques, in which the acquired diffraction patterns were binned and cropped around the central beam. In \resp{Figure~\ref{fig:nbcl_experiment}}(\resp{c}), in which diffraction patterns are binned by 3 and then cropped to $12\times12$ pixels, the dumbbells are still visible in both the class average and object itself. However, in \resp{Figure~\ref{fig:nbcl_experiment}}(\resp{d}), where the patterns are first binned by 6 and then the central $14\times14$ pixels were cropped, the dumbbells again are no longer resolvable, although like in \resp{\ref{fig:nbcl_experiment}}(a), their positions can be found. \resp{Figure~\ref{fig:nbcl_experiment}}(\resp{e}) shows the Fourier Transform of \resp{\ref{fig:nbcl_experiment}}(\resp{c}), with 6-fold symmetric reflections around the dotted circle indicating that the desired resolution to resolve the dumbbells was achieved.  In \resp{Figure~\ref{fig:nbcl_experiment}}(f), the Fourier Transform of \resp{\ref{fig:nbcl_experiment}}(\resp{d}), 6-fold symmetric spots are only observed corresponding to a maximum resolution of 0.84~\AA \resp{~when our reduced settings using sixfold binning are employed}. Both Fourier Transforms were averaged using a six-fold symmetry operation and weighted with a $r^{0.4}$ strength ramp to emphasize the signal from the noise.  As \resp{Figure~\ref{fig:nbcl_experiment}}(\resp{c} and) \resp{\ref{fig:nbcl_experiment}(e)} show, the potential for much faster data acquisition as well as reconstruction exists due to less pixels being required, when appropriately chosen acquisition settings are used.

\section{Conclusions}
\label{sec:conclusions}

In this paper we present the regularized optimization for ptychography (ROP) algorithm that uses derivative-based non-linear conjugate gradients optimization with Polak-Ribi\`ere search directions for the inversion.  Three error metrics were used:  the sum of absolute differences, the sum of squared differences and the negative log-likelihood for Poisson noise.

Through incorporation of the multislice formalism, multiple scattering was accounted for, resulting in an improved lateral resolution as demonstrated by its necessity in resolving the Nb-dumbbells in $4$~nm thick $\NbCl$ \resp{with and without data reduction}.

Data compromised by partially known experimental parameters, such as beam shape and positions, was successfully treated by optimizing said parameters along with the object.  Errors of up to $10$~nm in defocus and up to $0.024$~nm or $1.15$ times the step size in beam position were shown to be recoverable.

It was shown how the experimentally favorable settings of large step sizes and convergence angles, and small beam support and width of the recorded diffraction patterns, worsen the oversampling ratio and make the inverse problem less well-conditioned.  Regularization made reconstruction possible nonetheless, 
%even from severely underdetermined and noisy simulated data. 
\resp{even from noisy simulated data that was underdetermined by a factor of $0.47$.}
%Experimental data could also be significantly reduced while still obtaining the desired resolution.
\resp{The experimental data had its overdetermination  significantly reduced, from a factor of $70$ down to $5.6$, while still obtaining the desired resolution.}

These improvements relax the experimental requirements and hence are likely to extend the applicability of ptychography to thicker samples, and higher frame rates and scanning speeds, more convergent probes, and larger fields of view.  The reduced data load per recording furthermore facilitates a fast transfer of the experimental data and computation of the reconstruction.  This will help solidify the applicability of ptychography as a standard experimental technique for electron microscopy and optical setups, and increase its applicability to a broader class of microscopes and detectors.

\section*{Funding}

Deutsche Forschungsgemeinschaft (182087777 - SFB 951, BR 5095/2-1); Defense Advanced Research Projects Agency (TEE-D18AC00009); National Science Foundation (DMR-1539918, DMR-1719875).

\section*{Acknowledgments}
The authors thank I. El Baggari and L.F. Kourkoutis from Cornell University for the $\NbCl$ sample preparation.

Z.C. and D.A.M. are supported by PARADIM, a National Science Foundation (NSF) Materials Innovation Platform program. This work made use of the Cornell Center for Materials Research facility supported by NSF.

\paragraph{Additional Information.} Experimental data acquired at Cornell is available from D.A. Muller (david.a.muller@cornell.edu) on request. Code developed at Humboldt Universit\"at zu Berlin is available for scientific purposes from the first author (schlozma@hu-berlin.de) upon request.

\appendix

\section{Oversampling ratio for ptychography}
\label{app:underdata}

An expression for the oversampling ratio \cite{miao1998joptsoc} $N_\text{k}/N_\text{u}$ is derived. It depends on the experimental settings needed to attain a certain resolution $r$, corresponding to an angular frequency of $\thetareso$. 

% Under the phase object approximation 
\resp{When single scattering is dominant}
the angular frequency $\thetareso$ gets smeared out between $\thetareso - \thetaconv$ and $\thetareso + \thetaconv$. Attaining a resolution $r$ hence requires measuring diffraction patterns out to an angular frequency $\thetaobsv$ between these two values. For the same reason, in order for the forward model to accurately capture features of dimension $r$, the diffraction patterns need to be simulated out to at least $\thetacalc = \thetareso + \thetaconv$.  In the case of thin specimens, like the two-dimensional materials treated in this paper, negligible scattering is expected to exceed $\thetareso + \thetaconv$, and calculations beyond this limit are not needed.

The resolution $r$ is matched to the incoherent resolution, attained in high angle annular dark field scanning electron microscopy, by setting $\thetareso$ equal to $2\thetaconv$, and thus $\thetaconv < \thetaobsv \leq 3\thetaconv$.  The lower bound implies that ptychography is able to achieve the incoherent resolution limit from just the intensities in the central disc.

Within the calculated diffraction patterns, the region surrounding the cut-off disc, \textit{i.e.} the inner disc with diameter $2m/3$, is zeroed to avoid wrap-around artifacts \cite{kirklandBook}.  The angular frequency corresponding to the cut-off disc's edge is the maximum angle calculated and is denoted $\thetacalc$.  During the reconstruction process, the calculations are fitted to observations that extend out to  $\thetaobsv$, with generally $\thetaobsv  \leq \thetacalc$.  The width in pixels of the calculated diffraction patterns, $m$, is $3 \thetacalc / (\lambda \delta)$, and the equivalent width of the fitting region, $n$, is given by $2 \thetaobsv / (\lambda \delta)$.  The pixel width in reciprocal space, $\delta$, is $1/w$.  The pixel width in real space is $d = \lambda / (3 \thetacalc)$.

The number of knowns and unknowns, $N_\text{k}$ and $N_\text{u}$, is expressed as,
\begin{eqnarray*}
  N_\text{k} & = & n^2 \left( \frac{s}{\Delta x} + 1 \right)^2,\\
  N_\text{u} & = & 2 \left( \frac{ s + w }{d} \right)^2.
\end{eqnarray*}
The term of $1$ for $N_\text{k}$ prevents fencepost errors, and $\Delta x$ is the scan step.  The factor of $2$ in the expression for $N_\text{u}$ accounts for the complex nature of the reconstruction, and $s$ is the width of the scanned area. This yields,
\begin{eqnarray}
  \frac{N_\text{k}}{N_\text{u}} & = & \frac{2}{9} \frac{ \thetaobsv^2 }{ \thetacalc^2 } \frac{ w^2 }{ \Delta x^2 } \left( \frac{ 1 +  \Delta x / s  }{  1 + w/s  } \right)^2, \label{eq:NuNk} \\
  & \xrightarrow[s\to\infty]{} & \frac{2}{9} \frac{ \thetaobsv^2 }{ \thetacalc^2 } \frac{ w^2 }{ \Delta x^2 }, \label{eq:NuNkwidefield}
\end{eqnarray}
with the second equation describing a typical wide-field scan.  For thin specimens and the incoherent resolution limit, $\thetacalc$ is substituted with $3 \thetaconv$ and (\ref{eq:NuNkwidefield}) reduces further to~(\ref{eq:NuNkRed}). 

\resp{

\subsection{Guidelines}

From~(\ref{eq:NuNkRed}) guidelines for experimental settings are deduced. Expressing that $\Nk/\Nu > 1$ leads to 
\begin{align}
    w & > 2.1 \Delta x, \text{ for } \thetaobsv = 3 \thetaconv, \text{ and}\\
    w & > 6.4 \Delta x, \text{ for } \thetaobsv = \thetaconv.
\end{align}
The approximate beam width is given by the Rayleigh criterion as  $0.61 \lambda / \thetaconv$, and practical experience with both the experimental and simulated data sets in this paper, has indicated that $\Delta x$ should be considerably lower than this value, for example two-thirds or a half.  

To ensure an acceptable forward simulation, $w$ must be wide enough to encompass a sizeable fraction of the beam intensity.   At $w \simeq (k+0.24) \lambda / \thetaconv$ the support encloses the first $k$ minima of a focused diffraction limited probe.  It is hence recommended to choose $w$ of the order $2 \lambda / \thetaconv$ or above.

}

\section{Derivatives}
\label{app:derivatives}

Here, the derivative of the loss function $\ell$ with respect to the incoming beam shape and beam positions as given by Eqs.~(\ref{eq:DlDPsiRe}), (\ref{eq:DlDPsiIm})~and~(\ref{eq:DlDR}) are derived.

For the incoming beam $\psi^0$, which is a complex quantity, the derivative can be expressed for its real $\psi^0_{re}$ and imaginary part $\psi^0_{im}$, respectively. According to the chain rule for Wirtinger derivatives \cite{wirtinger1927formalen}, we have:
\begin{equation}
    \frac{ \partial \ell }{\partial \psi^0_{re}} = \frac{\partial \ell}{\partial \psi^0 t^0} \frac{\partial \psi^0 t^0}{\partial \psi^0_{re}} + \frac{\partial \ell}{\partial (\psi^0 t^0)^{*}} \frac{\partial (\psi^0 t^0)^{*}}{\partial \psi^0_{re}}, 
\end{equation}
where $*$ being the complex conjugate, $\partial \ell / \partial \psi^0 t^0$ obtainted from the backpropagation and $\partial \ell / \partial (\psi^0 t^0)^{*} = ( \partial \ell / \partial \psi^0 t^0)^{*}$. The equivalent expression for the imaginary part is omitted. Equation~(\ref{eq:DlDPsiRe}) then follows from $\partial \psi^0 t^0 / \partial \psi^0_{re} = t^{0}$ and $\partial (\psi^0 t^0)^{*} / \partial \psi^0_{re} = t^{0*}$ and equation~(\ref{eq:DlDPsiIm}) follows from $\partial \psi^0 t^0 / \partial \psi^0_{im} = it^{0}$ and $\partial (\psi^0 t^0)^{*} / \partial \psi^0_{im} = it^{0*}$.

For the derivative of the beam positions we keep the notation light, by only treating the one-dimensional case, the extension to two dimensions being trivial. The first element of $\vecr$ and  $\vecR$ are denoted $x$ and $X$, respectively. The probe positioned in the origin is denoted as $\psi^0$, and in position $X$ as $\psi^X$, with $\psi^X(x) = \psi^0( x - X )$. Following the chain rule:
\begin{equation}
    \frac{ \partial \ell }{\partial X} = \sum_k \frac{\partial \ell}{\partial \psi^X_k} \frac{\partial \psi_k^X}{\partial X} + \frac{\partial \ell}{ \partial \psi^{X*}_k} \frac{\partial \psi_k^{X*}}{\partial X}, \label{eq:charul}
\end{equation}
with $\partial \ell / \partial \psi_k^X$ provided by the backpropagation. Equation~(\ref{eq:DlDR}) is then obtained with $\partial \psi^X / \partial X = -\partial \psi^X / \partial x$ and $\partial \ell / \partial \psi_k^{X*} = ( \partial \ell / \partial \psi_k^{X} )^*$.

\resp{
\section{Algorithm implementation}

All the reconstructions presented were generated with a ROP implementation that utilizes the parallel hardware architecture of a NVIDIA V100 GPU, allowing a simultaneous propagation through the network for a batch of probe positions. For a batch size of $100$, convergence of the algorithm for the tasks described in section 3.1 ranged between $3$ to $4\text{h}$ and for every task in section 3.2 roughly $5\text{min}$. Experimental results in section 3.3 were obtained with a batch size of $87$ for the reconstructions of the $\MoS$ data and a batch size of $10$ for the reconstructions of the $\NbCl$ data, with an average time of $15\text{min}$ and $40\text{min}$, respectively.}

\section*{}

\begin{table*}[htbp]
\caption{Settings for simulation, and reconstruction from simulated and experimental measurements of $\text{Mo}\text{S}_2$ and $\text{Nb}_3\text{Cl}_8$. The acceleration voltage is $80$~kV ($\lambda = 4.18$~pm) and $120$~kV ($\lambda = 3.35$~pm), respectively.  To treat the non-square scan patterns, the width $s$ of the scan area is approximated as the width of a square of equal area. $N_\text{k}/N_\text{u}$ has been calculated from (\ref{eq:NuNk}).}
\label{tab:setupNbCl}
\center
\begin{tabular}{l | c c c | c c c c c c }
  \hline
  \hline 
  & \multicolumn{3}{c|}{$\text{Mo}\text{S}_2$} & \multicolumn{6}{c}{$\text{Nb}_3\text{Cl}_8$}\\
  					 & \scriptsize{Sim} 			& \scriptsize{Sim Rec} 	& \scriptsize{Exp Rec} 	& \scriptsize{Grnd Trth} 	& \scriptsize{Sim} 		& \scriptsize{Sim Rec} 	& \scriptsize{Exp Rec 1} 	& \scriptsize{Exp Rec 2} 	& \scriptsize{Exp Rec 3}  \\
  \hline
  $\theta_{\text{con}}$ (mrad)                    & $21.4$		& $21.4$ 	& $21.4$     	& $28$ 		& $28$ 		& $28$ 		& $24$ 		& $24$ 		& $24$ \\
  $\thetareso$ (mrad)                    & --- 			& --- 	& ---        & $56$ 		& --- 		& $56$ 		& $95.0$ 	& $55.8$ 	& $55.8$ \\
  $m$                                    & $256$ 		& $256$ 	& $264$      	& $40$ 		& $80$ 		& $20$ 		& $264$ 	& $40$ 		& $22$ \\
  $n$                                    & $124$ 		& $124$ 	& $124$      	& $8$ 		& $16$ 		& $4$ 		& $124$ 	& $12$ 		& $14$ \\
%   $d$ (nm)                               & $0.0118$ 		& $0.0118$ 	& $0.0115$  	& $0.0116$ 	& $0.0116$ 	& $0.0116$ 	& $0.00661$ 	& $0.0143$ 	& $0.0130$ \\
$d$ (pm)                               & $11.8$ 		& $11.8$ 	& $11.5$  	& $11.6$ 	& $11.6$ 	& $11.6$ 	& $6.61$ 	& $14.3$ 	& $13.0$ \\
  $\delta$ $\left(\text{nm}^{-1}\right)$ & $0.331$ 		& $0.331$ 	& $0.331$   	& $2.15$ 	& $1.08$ 	& $4.31$ 	& $0.573$ 	& $1.75$ 	& $3.51$ \\
%   $\Delta x$ (nm)                        & $0.021$ 		& $0.021$ 	& $0.021$    	& $0.0229$ 	& $0.0457$ 	& $0.0457$ 	& $0.043$ 	& $0.043$ 	& $0.043$ \\
$\Delta x$ (pm)                        & $21.0$ 		& $21.0$ 	& $21.0$    	& $22.9$ 	& $45.7$ 	& $45.7$ 	& $43.0$ 	& $43.0$ 	& $43.0$ \\
  $s$ (nm) 				 & $2.10$ 		& $1.40$ 	& $1.40$ 	& $3.71$ 	& $3.71$ 	& $3.71$ 	& $2.75$ 	& $2.75$ 	& $2.75$ \\
  $N_\text{k}/N_\text{u}$                              & $4.2 \scE 2$ 	& $2.5\scE 2$ 	& $2.4 \scE 2$  & $6.6$ 	& $5.4$ 	& $0.47$ 	& $7.0 \scE 1$ 	& $5.6$ 	& $7.5$ \\
  \hline
  \hline
\end{tabular}
\end{table*}

\begin{table}[htbp]
\caption{Additional ROP settings for the $\text{Mo}\text{S}_2$ experimental reconstructions, corresponding to panels in Figure~\ref{fig:mos2_experiment}.}
\label{tab:setupmosexp}
\center
\begin{tabular}{ c | c c c c }
  \hline
  \hline
  & (a) & (b) & (c) & (d) \\
  \hline
  $\mu$ & 0.0215 & 0 & 0.01 & 0.01 \\
  $\alpha_{0}$ & 1 & 1 & 1 & 1 \\
  $\beta_{0}$ & -- & $1 \scEm 3$ & $1 \scEm 3$ & $1 \scEm 3$ \\
%   $\gamma_{\text{init}}$ & -- & -- & -- & -- \\
  Sub ep. $\alpha$ (iter.) & 3 & 2 & 3 & 3 \\
  Sub ep. $\beta$ (iter.) & -- & 10 & 15 & 15 \\
%   Iter.$_\gamma$ & -- & -- & -- & -- \\
%   \# of slices & 1 & 1 & 1 & 1 \\
  Err. Func. & $\mathcal{E}_3$ & $\mathcal{E}_1$ & $\mathcal{E}_3$ & $\mathcal{E}_3$ \\
  \hline
  \hline
\end{tabular}
\end{table}

\begin{table}[htbp]
\caption{Additional ROP settings for the $\text{Nb}_3\text{Cl}_8$ experimental reconstructions, corresponding to panels in Figure~\ref{fig:nbcl_experiment}.}
\label{tab:setupnbclexp}
\center
\begin{tabular}{ c | c c c c }
  \hline
  \hline
  & (a) & (b) & (d) & (e) \\
  \hline
  $\mu$ & 1\scE-6 & 10 & 0.1 & 20 \\
  $\alpha_{0}$ & 5 & 5 & 5 & 5 \\
  $\beta_{0}$ & $5\scEm5$ & $1\scEm4$ & $5\scEm5$ & $1\scEm4$ \\
%   $\gamma_{\text{init}}$ & -- & -- & -- & -- \\
  Sub ep. $\alpha$ (iter.) & 10 & 10 & 10 & 10 \\
  Sub ep. $\beta$ (iter.) & 3 & 3 & 3 & 3 \\
%   Iter.$_\gamma$ & -- & -- & -- & -- \\
  \# slices & 1 & 6 & 6 & 6 \\
  $\Delta z_{\text{slice}}$ (nm) & -- & 0.67 & 0.67 & 0.67 \\
  Err. Func. & $\mathcal{E}_3$ & $\mathcal{E}_1$ & $\mathcal{E}_3$ & $\mathcal{E}_1$ \\
  \hline
  \hline
\end{tabular}
\end{table}

%%%%%%%%%%%%%%%%%%%%%%% References %%%%%%%%%%%%%%%%%%%%%%%%%

\FloatBarrier

\bibliographystyle{unsrt}
\bibliography{ROP-References}

% \printbibliography

\end{document}